\documentclass[prl,aps,twocolumn,superscriptaddress,showpacs]{revtex4}

\usepackage[tbtags]{amsmath}
\usepackage{amssymb}
\usepackage{amsfonts}
\usepackage[dvips]{graphicx,color}
\usepackage{amsmath}
\usepackage{amssymb}
\usepackage[dvips]{graphicx}
\usepackage{graphicx}
\usepackage{dcolumn}
\usepackage{bm}
\usepackage{setspace}
\usepackage{amsfonts,color}
\usepackage{rotating}
\usepackage{makeidx}
\usepackage{amsthm}
\usepackage{xcolor}

\usepackage{url}
\newcommand{\Tr}{\mathrm{Tr}}

\usepackage{appendix}

\usepackage[colorlinks=true,citecolor=blue,urlcolor=blue]{hyperref}

\begin{document}

\title{ Steerability detection of arbitrary 2-qubit state via machine learning}
\author{Changliang Ren}\thanks{These authors contributed equally.}
\affiliation{Center for Nanofabrication and System Integration, Chongqing Institute of Green and Intelligent Technology, Chinese Academy of Sciences, Chongqing 400714, People's Republic of China}\email{ renchangliang@cigit.ac.cn}\affiliation{CAS Key Laboratory of Quantum Information, University of Science and Technology of China, Hefei 230026, PR China}

\author{Changbo Chen}\thanks{These authors contributed equally.}
 \affiliation{Chongqing Key Laboratory of Automated Reasoning and Cognition, Chongqing Institute of Green and Intelligent Technology, Chinese Academy of Sciences, Chongqing 400714, People's Republic of China}\email{chenchangbo@cigit.ac.cn}

%
%

\date{\today}
\begin{abstract}
 Quantum steering is an important nonclassical resource for quantum information processing. However, even lots of steering criteria exist, it is still very difficult to efficiently determine whether an arbitrary two-qubit state shared by Alice and Bob is steerable or not, because the optimal
 measurement directions of Alice are unknown. In this work, we provide an efficient quantum steering detection scheme for arbitrary 2-qubit states with the help of machine learning, where Alice and Bob only need to measure in a few fixed measurement directions. In order to prove the validity of this method, we firstly realize a high performance quantum steering classifier with the whole information. Furthermore, a high performance quantum steering classifier with partial information is realized, where Alice and Bob only need to measure in three fixed measurement directions. Our method outperforms the existing methods in generic cases in terms of both speed and accuracy, opening up the avenues to explore quantum steering via the machine learning approach.
\end{abstract}


\maketitle

\section{I. Introduction}

In the great debate of quantum mechanics in 1930s, Einstein, Podolsky, and Rosen (EPR) \cite{Einstein} challenged the completeness of quantum mechanics (QM) based on local realism usually called EPR paradox. It points out a way to deeply investigate the difference or conflict between classical theory and quantum theory. Especially, three types of quantum correlations originated from EPR paradox: quantum entanglement \cite{Horodecki}, Bell nonlocality \cite{Brunner}, and EPR steering \cite{Cavalcanti}, have been put forward. Within the hierarchy of nonlocalities, the set of EPR steerable states is a subset of entangled states and a superset of Bell nonlocal states. Quantum entanglement and Bell nonlocality have attained flourishing developments since 1964. However, a rigorous formulation of the concept of EPR steering was not elaborately interpreted until 2007 \cite{Wiseman}. Recently, it has gained increasing interest in quantum optics and quantum information communities \cite{Wiseman,Jones,Skrzypczyk}. For instance, EPR steering can provide security in one-sided device-independent quantum key distribution (1SDI-QKD) \cite{Branciard,Gehring,Walk} and play an operative role in channel discrimination \cite{Piani} and
teleamplification \cite{He}.

 Naturally, detection and characterization of steering have attracted increasing attention \cite{Wiseman,Reid,Reid1,Jones,Piani,Cavalcanti1,Cavalcanti2,Walborn,Schneeloch,Pusey,Pramanik,Kogias,Skrzypczyk,Kogias1,Zhu,Nguyen}. Various steering criteria and inequalities have been derived, such as linear steering inequalities \cite{Cavalcanti1,Cavalcanti2,Pusey}, inequalities based on multiplicative variances \cite{Reid,Reid1,Cavalcanti1}, entropic uncertainty relations \cite{Walborn,Schneeloch}, fine-grained uncertainty relations \cite{Pramanik}, and hierarchy of steering criteria based on moments \cite{Kogias}. In particular, a necessary and sufficient condition for a two-qubit state to be steerable with respect to projective measurements is exhibited \cite{Nguyen}. Actually, these criteria can be computed through semidefinite programming \cite{Cavalcanti3}.

 For an arbitrary quantum state shared by Alice and Bob, to determine if Alice can steer Bob, those criteria boils down to finding optimal measurement directions of Alice, which is resource demanding as explained below. In real experimental test, if Alice and Bob share an unknown state, there are two ways to identify the steerability of this state. One is computing through the steering criteria after a complete quantum state tomograph measurement, the other is trying to directly observe the characterized phenomenon (such as the violation of the steering inequality),  which can distinguish the steerability and non-steerability. Obviously, the former needs to measure the whole information of the state, while the latter has to try many times by choosing a large amount of measurement directions until the characterized phenomenon is observed. Hence, both of them need a lot of measurements and not efficient. It is even worse when there are a large amount of different states to be detected, which is typical if one wants to detect steerability of a sequence of distinct rapidly generated  states.
 Thus it remains challenging to develop an efficient approach to detect steerability for experimental test.


 %
%

Recently, the successful applications of machine learning approach on entanglement \cite{Lu,Ma} and nonlocality  discriminants \cite{Deng} shed a new light on this problem.
Machine learning
possesses the capability to instantly make predictions on new data with reasonable accuracy
after learning from large amount of existing data. In the past few decades, there has been a rapid growing interest not only in theoretical studies,
but also in a variety of applications of machine learning.
Interestingly, beside its extensive applications in industry,
machine learning has also been employed to investigate physics-related problems in recent years.
Nowadays, many quantum implementations of machine learning have been introduced to achieve better performance for quantum information
processing \cite{Deng,Lu,Ma,Assion,Sasaki,Bisio,Hentschel,Bang,Wiebe,Krenn,Schoenholz,Zhang}, such as the hamiltonian learning \cite{Wiebe}, automated quantum experiments search \cite{Krenn}, phase transition identification \cite{Schoenholz}, identification of topological phase of matter \cite{Zhang}, entanglement classification \cite{Deng,Lu,Ma}, just to name a few.

Certainly, these works motivate us to adopt machine learning as an alternative approach for investigations of various quantum tasks. Different from the previous researches, in this paper, we employ the machine learning techniques to tackle the bipartite steering detection problem by recasting it as a learning task. We build several new steerability classifiers underpinned by machine learning techniques. Firstly, an efficient steerability classifiers with the whole information demonstrated the validity of steering classification by machine learning. Secondly, and more importantly, we provide a quantum
steering detection scheme for arbitrary two-qubit states with the help of machine learning, where
Alice and Bob only need to measure in three fixed measurement directions.
These efficient steerability classifiers, which work for arbitrary 2-qubit states, are exhibited and fully analyzed.
Either for arbitrary 2-qubit state or special states, they can perform better than the traditional semidefinite programming (SDP).
More importantly, comparing with the classical method, this approach is much less resource demanding
and can quickly determine whether a state is steerable with a well-trained classifier.
Hence, it provides a simpler and more efficient way to detect steerability, 
which sheds new light on classification of quantum steering with limited resources, 
and represents a step towards large-scale machine-learning-based applications in quantum information processing.

\section{II. Quantum steering}

We start by defining the scenario in which quantum steering
is discussed. For the sake of convenience, let us only consider the simplest case --- two qubit system. Consider a bipartite situation composed by Alice and Bob sharing an arbitrary quantum state $\rho$. Suppose Alice performs measurement $\hat{A}$ with outcome $a$ and Bob performs measurement $\hat{B}$ with outcome $b$. These outcomes are thus in general governed by a joint probability distribution $P(a,b\mid\hat{A},\hat{B},\rho)$. Such joint probability distribution predicted by quantum theory is defined by
\begin{eqnarray}\label{QM}
P(a,b\mid\hat{A},\hat{B},\rho)=\mathrm{Tr}(\hat{M}^{a}_{A}\otimes \hat{M}^{b}_{B}\rho),
\end{eqnarray}
where $\hat{M}^{a}_{A}$ and $\hat{M}^{b}_{B}$ are the projective operators for Alice and Bob respectively.

It is well-known that, Wiseman, Jones and Doherty formally defined quantum steering as the possibility of remotely generating ensembles
that could not be produced by a local hidden state (LHS) model. The mathematic formulation of the LHS model adds an extra requirement on Bob's probabilities, which can be expressed as
\begin{equation}\label{LHS}
\begin{array}{rcl}
P(a,b\mid\hat{A},\hat{B},\rho)&=&\sum_{\lambda}P(\lambda)P(a\mid\hat{A},\lambda)P_{Q}(b\mid\hat{B},\lambda)\quad\\
P_{Q}(b\mid\hat{B},\lambda)&=&\mathrm{Tr}(\rho_{\lambda}\hat{M}^{b}_{B}),
\end{array}
\end{equation}
where $\rho_{\lambda}$ is a qubit specified by $\lambda$. If the joint probability can be decomposed in the form of Eq. (\ref{LHS}), then we say that Alice can not steer Bob's state.
Otherwise $P(a,b\mid\hat{A},\hat{B},\rho)$ shows quantum steering correlation (in the sense that Alice steers Bob). The steering scenario consists of the situation where no
characterisation of Alice's measurements is assumed, while Bob has full control of his measurements and can thus access the unnormalised conditional states $\sigma_{a\mid A}$, where $\sigma_{a\mid A}=\mathrm{Tr}_A[(\hat{M}^{a}_{A}\otimes I) \rho]$. In other words, deciding whether an assemblage $\sigma_{a\mid A}$
demonstrates steering amounts to checking whether there exists
a collection of quantum states $\rho_{\lambda}$ and probability distributions
$P(\lambda)$ and $P(a\mid\hat{A},\lambda)$ such that (\ref{LHS}) holds.
Obviously this is in
principle a hard problem, since the variable $\lambda$ could assume
infinitely many values. However, if the
number of measurements and outputs is finite, this problem becomes
much simpler, and it was shown in \cite{Cavalcanti} that
the problem can be solved through
semi definite programming (SDP) \cite{Vandenberghe}.
Next, we briefly review this approach.

Suppose that Alice performs $m$ measurements, labeled as $x=0,1......,m-1$.
One can write a SDP that determines if Alice can steer Bob \cite{Cavalcanti},
\begin{equation}
\label{eq1}
\begin{array}{lll}
&\mbox{given}                       & \{\sigma_{a|x}\}, \{D(a|x, \lambda')\}_{\lambda'}\\
&\underset{\{F_{a|x}\}}{\mbox{min}} & \Tr\sum_{ax} F_{a|x}\sigma_{a|x}\\
&\mbox{s.t.}                        &\sum_{ax}F_{a|x}D(a|x,\lambda')\geq 0\;\;\;\forall \lambda'\\
&                                   &\Tr\sum_{ax,\lambda'} F_{a|x}D(a|x,\lambda')=1,
\end{array}
\end{equation}
where $F_{a|x}$ are Hermitian matrices, $\lambda'$ is a map from $\{x\}$ to $\{a\}$ and
$D(a|x,\lambda')=\delta_{a,\lambda'(x)}$, that is $D(a|x,\lambda')=1$ if $\lambda'(x)=a$ and $D(a|x,\lambda')=0$ if $\lambda'(x)\neq a$.
If the objective value of~(\ref{eq1}) is negative for some measurements $x$, then $\rho$ is steerable from Alice to Bob. On the other
hand, a non-negative value means that there exists an LHS model.

%


\section{III. Quantum steering classifier with whole information}

Theoretically, we can detect steering more and more precise with the increase of measurements by SDP \cite{Cavalcanti3}. However, there is yet
a noticeable drawback of the above SDP approach from the
perspective of the tradeoff between the accuracy and time consumption.
Boosting the accuracy means adding additional extreme
points to enlarge the convex hull, which requires more time to determine if a point is inside the enlarged convex
hull or not. To overcome this, we combined SDP with supervised
learning, as machine learning has the power to speed up
such computations.

Naturally, the bipartite steering detection
problem can be formulated as a supervised binary classification
task. Here, the datasets are generated by adopting the following procedure:
\begin{itemize}
\item First generating two random $4\times 4$ matrices $M$ and $N$, which are used to
generate a Hermitian matrix $H := (M+iN)(M+iN)^{\dag}$, where $\dag$ means taking the conjugate transpose, and a density matrix $\rho_{AB} := H/{\rm Tr}(H)$.
\item Since $\rho_{AB}$ is a density matrix of $4\times 4$, it is enough to use the first $3$ elements on the diagonal
and the real and imaginary parts of $6$ elements below the diagonal of the matrix to form
the vector of features, which is a real vector of $15$ numbers in the interval $(-1, 1)$, denoted by $F_1$.
\item For a given density matrix $\rho_{AB}$, we run SDP Program~(\ref{eq1}) $100$ times with different
values of measurements. If the objective value is negative, we assign a label $-1$;
  otherwise we assign a label $+1$, which means that we do not know if Alice can steer Bob.
\end{itemize}
For each $m=2,\ldots,8$, we generate the corresponding dataset until at least $5000$ samples with label $+1$ and $5000$ samples with label $-1$
are obtained.
Generating the datasets for all different settings $m=2,\ldots,8$ take several months on a workstation.
Here, we should emphasize that the collected states for different settings $m$ are totally random and independent.
Finally, we collected over $70000$ samples in total \cite{Chen}. Actually, it becomes
harder and harder to obtain a dataset with the increase of amount of measurements. For example, when $m=8$, it spent about $63$ days to collect the dataset.
For each $m=2,\ldots,8$, the last $1000$ positive samples and the last $1000$ negative samples are reserved for test.
The rest $4000$ positive samples and $4000$ negative samples are kept as the training set to learn a classifier.
We employ a $4$-fold cross-validation technique and a grid search approach for selecting best hyperparameters.
The machine learning method we use is support vector machine (SVM).

SVM is a supervised learning model used for classification and regression analysis, which requires solving the following optimization problem:
\begin{equation}
\label{svm}
\begin{array}{lll}
&\mbox{given}                       & ({\bf x}_i, y_i), i=1,\ldots,\ell \\
&\underset{{\bf w}, b, \mathbf{\xi}}{\mbox{min}} & \frac{1}{2}{\bf w}^T{\bf w}+C\sum_{i=1}^{\ell}\xi_i\\
&\mbox{s.t.}                        &y_i({\bf w}^T\phi({\bf x}_i)+b)\geq 1-\xi_i\\
&                                   &{\xi}_i\geq 0.
\end{array}
\end{equation}
Here $\ell$ is the number of samples, $y_i$ and ${\bf x}_i$ are respectively
the label and the vector of features of sample $i$, $\phi$ is a mapping implictly defined by a kernel
function and  we choose the radial basis function (RBF) kernel $K(\phi({\bf x}_i)^T, \phi({\bf x}_j))={\rm exp}(-\gamma\lVert{{\bf x}_i-{\bf x}_j}\rVert^2)$
with the parameters $C$ and $\gamma$ to be determined by a grid search approach when training the model.

\begin{figure}
\begin{center}
\includegraphics[width=0.45\textwidth]{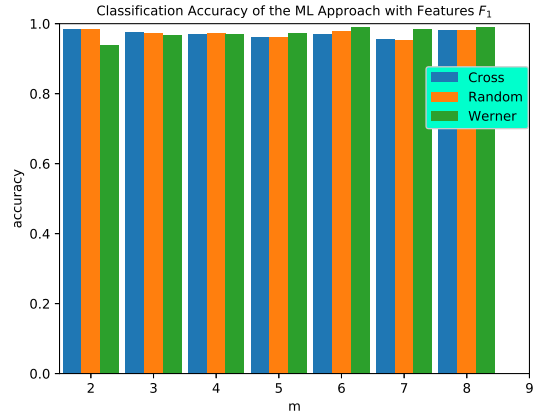}
 \caption{\label{fig:accuracy}(Color online) \small{Classification accuracy of machine learning with the whole information ($F_1$-features). The first column (blue) depicts the accuracy of cross validation; the second column (orange) depicts the classification accuracy on random states and the third column (green) depicts the classification accuracy on the Werner state.}}
\end{center}
\end{figure}

In the rest of this section, the models are trained with feature vectors of type $F_1$, which encodes
the full information of a two-qubit quantum state.
After the SVM model is trained, we test the performance by creating a new set of quantum ensemble that is
distinct from the data set employed for training.
The classification accuracy of the learning model for each $m$ is illustrated in Fig.~\ref{fig:accuracy}.
All the accuracies for training and cross validation are higher than $0.95$, which clearly shows that the models are well-trained.

\begin{figure}
\begin{center}
\includegraphics[width=0.45\textwidth]{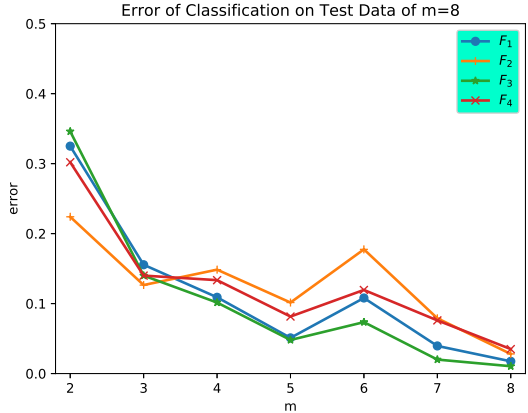}
\caption{\label{fig:test8}(Color online) \small{Error of classification on test data of $m=8$ by each machine learning classifier
    with different features $F_1, F_2,F_3,F_4$ (summarized in appendix) respectively. }}
\end{center}
\end{figure}

It is reasonable to predict that, if these steerability classifiers are well-trained, the classifiers should turn more precise with the increase of $m$.
To show such validity, we use classifiers learned for different $m$ to test against
the random data for $m=8$. As illustrated in Fig.~\ref{fig:test8}, the blue-circle line is for $F_1$ features,
it is shown that, the error drops very rapidly (except $m=6$, which may come from the imperfection in the learning process).
However, the variation tendency is identical to the theoretical prediction in general (the more measurement settings, the more precise the prediction)
which further demonstrates that these are well-trained classifiers of quantum steering.

To demonstrate the generalization ability of the steering classifiers,
we study their ability in clarifying a special state which has unambiguous bound for steerability.
The state can be written as,
\begin{eqnarray}\label{target state}
\rho_{\mathrm{W}} &=& p\mid\psi\rangle\langle\psi\mid + (1-p)\rho_{\mathrm{A}} \otimes I/2
\end{eqnarray}
where $\mid\psi\rangle =\cos \xi \mid 00\rangle + \sin \xi \mid 11\rangle$, $\rho_{\mathrm{A}}=\mathrm{Tr_B}(\mid\psi\rangle\langle\psi\mid )$. This state is a two-qubit one-way steerable
state, which was exhibited by \emph{Bowles et.al} in \cite{Bowles} recently. The state reduces to the Werner state when $\xi=\frac{\pi}{4}$. In simplicity, this state can be called ``generalized Werner state". Different from entanglement and nonlocality, each qubit in Alice and Bob plays different role in the steering scenario. There exists one-way quantum steering. That is, special entangled states such that steering can occur from Alice to Bob, but not from Bob to Alice. One-way steering states attracted more attention due to their special characterization. For the state in Eq. (\ref{target state}), which is unsteerable from Alice to Bob \cite{Bowles}, if
\begin{eqnarray}\label{one-way}
\cos^{2}2\xi\geq\frac{2p-1}{(2-p)p^3}
\end{eqnarray}
which has also been experimentally demonstrated very recently in \cite{Xiao}.
Obviously, the bound of the parameter $p$ that Alice can steer Bob's state is determined by Eq. (\ref{one-way}).
Here we apply our classifiers to predict the steerability of such states.
The test states are constructed according to the uniform distribution of $p$ and $\xi$.
For each $\xi=\{\frac{\pi}{4},\frac{\pi}{6},\frac{\pi}{8},\frac{\pi}{12}\}$,
we generate $10000$ test samples.
We predict the steerability bounds using both learned classifiers and SDP.

\begin{figure}[h]
\begin{center}
\includegraphics[width=0.45\textwidth]{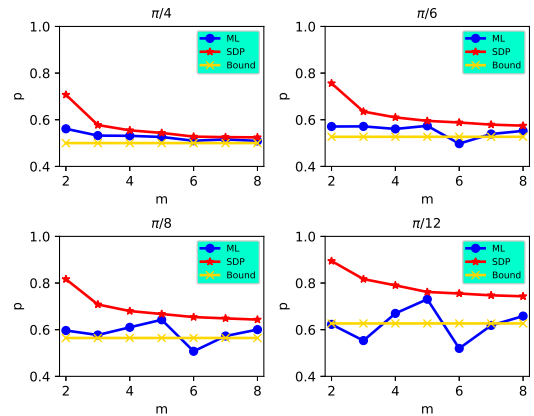}
 \caption{\label{fig:angle}(Color online) \small{The predictions of steerability for generalized Werner states by learned classifiers and SDP. The blue line is the result  predicted by learning classifiers with $F_1$ features. Similarly, the red line is the result  predicted by SDP. And the orange line is the steerability bound from Alice to Bob.}}
 \end{center}
\end{figure}

\begin{figure}[h]
\begin{center}
\includegraphics[width=0.45\textwidth]{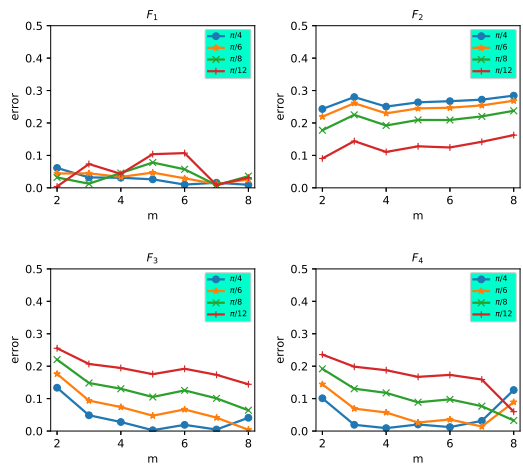}
 \caption{\label{fig:error-Fs}(Color online) \small{Classification error for generalized Werner states with different classifiers.}}
\end{center}
\end{figure}

As illustrated in Fig.~\ref{fig:angle}, four subfigs in this picture correspond to $\xi=\{\frac{\pi}{4},\frac{\pi}{6},\frac{\pi}{8},\frac{\pi}{12}\}$ respectively.
In each subfig, the blue line is the result predicted by the learned classifiers for $m=2,...,8$ respectively.
Similarly, the red line is the result predicted by SDP with $m=2,...,8$.
The orange line is the steerability bound from Alice to Bob which is defined by Eq. (\ref{one-way}).
Obviously, the learning classifiers perform better than the traditional SDP.
Especially, when $\xi=\frac{\pi}{4}$, Werner state, the learning classifiers demonstrate the best performance.
In Fig.~\ref{fig:accuracy}, the third green column depicts the classification accuracy on the Werner state.
It is interesting to notice that for $m>4$, the classification accuracy on Werner state is even higher than on random data,
despite the fact that the model is trained with random data.
As the decrease of $\xi$, the prediction errors of both the learning classifiers and SDP increase.
It is a reasonable phenomenon since the predictions for the marginal states become harder and harder.
In Fig.~\ref{fig:error-Fs}, the first subfig for $F_1$ features shows the classification error for generalized Werner States (with different angles).
Obviously, the error increases when the angle drops which coincides with the above analysis.
Another interesting phenomenon is, even though the learned classifiers can be more effective than SDP,
it is still possible that they may predict the value of $p$ lower than the steerability bound,
which almost never happens for SDP. The reason is that, one typical character for machine learning classifiers is that
they can predict the positive to be negative and vice versa.
However, the main error for SDP occurs when it predicts the negative to be positive.
Hence, this phenomenon can be used to distinguish which method is used.

The above results clearly demonstrate the validity of steerability detection by machine learning. 
Even the whole information is still needed by this scheme, same as the traditional SDP method, 
the machine learning method is much more efficient than SDP in data processing.
Take $m=8$ for example, the learned classifier spends about $10^{-2}\mathrm{s}$ to predict an unknown state while
it takes about $10^2\mathrm{s}$ for the SDP with $m=8$.
Maybe it is unfair to exhibit the time advantage in testing only one state,
after all the time cost of the machine learning classifier should contain both the training time and the prediction time.
However, when the task is to predict a large number of unknown states,
the time advantage of machine learning classifier is obvious.

\section{IV. Quantum steering classifiers with partial information}

Although the above steering classifier via machine learning boosts the performance of the state classification compared with traditional SDP method,
it still has a disadvantage that such classifier needs the whole information of the state as the input feature. 
However, the size of a quantum state grows exponentially when scaled up, which makes large-scale quantum state
tomography intractable to carry out. Hence, it will become more and more difficult to extend
the method to higher dimensions. 
Hence, it is important to further explore the possibility of learning with only partial information of the quantum state. 
Here we will introduce an efficient quantum steering detection scheme for arbitrary two-qubit states with the help of machine learning, 
where Alice and Bob only need to measure in a few fixed measurement directions.

Steerability is unaffected by local unitaries for Alice and ``local filters"/''stochastic local operations" for Bob. Hence the relevant information for steerability could be encoded in a smaller feature vector. Actually, an arbitrary two-qubit state can be expressed in the local Pauli basis
{\small
\begin{eqnarray}\label{state}
\rho= \frac{1}{4} \biggr(I + \sum_{i=1}^{3}r_{i}\sigma_{i} \otimes I  +
\sum_{j=1}^{3} s_{j}I\otimes \sigma_j + \sum_{k,l=1}^3 \tau_{kl} \sigma_k
\otimes \sigma_l \biggr).
\end{eqnarray}
}
Steerability is determined by all the parameters $r_{i}, s_{j},\tau_{kl}$. It is intuitively believed that steerability is dominated by the correlation terms $\tau_{kl}$ between the two qubits from Eq. (\ref{state}). Hence, it is natural to extract the coefficients of the correlation terms, $ \{\tau_{kl}\}$, as features.
More precisely, the partial information is extracted by computing $\mathrm{Tr}[(\sigma_k\otimes \sigma_l)\rho]$ as features, denoted by $F_2$.

\begin{figure}
\begin{center}
\includegraphics[width=0.45\textwidth]{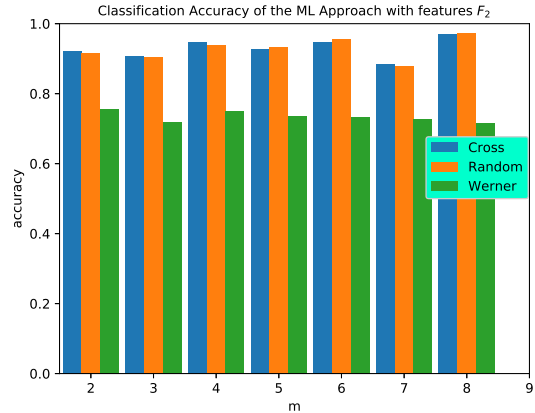}
 \caption{\label{fig:accuracy-F2}(Color online) \small{Classification accuracy of machine learning with partial information ($F_2$ features). The first column (blue) depicts the accuracy of cross validation; the second column (orange) depicts the classification accuracy on random states and the third column (green) depicts the classification accuracy on the Werner state.}}
\end{center}
\end{figure}

\begin{figure}[h]
\begin{center}
\includegraphics[width=0.45\textwidth]{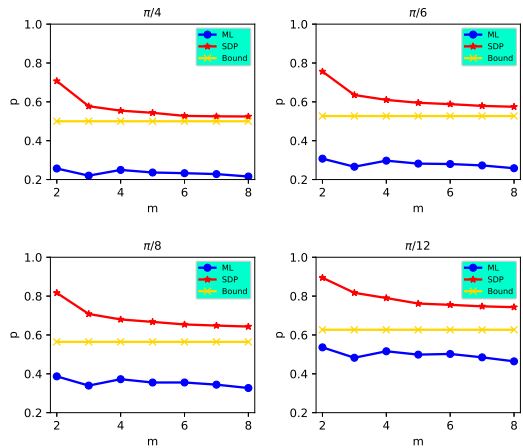}
 \caption{\label{fig:errorBound-F2}(Color online) \small{The predictions of steerability for generalized Werner states by learning classifiers and SDP. The blue line is the result which predicted by learning classifiers with $F_2$ features. Similarly, the red line is the result which predicted by SDP. And the orange line is the steerability bound from Alice to Bob.}}
 \end{center}
\end{figure}

We repeat the same training and test process as for $F_1$.
The classification accuracy of the learned models for each $m$ is illustrated in Fig.~\ref{fig:accuracy-F2}.
It is interesting to see that, even the classification accuracy on random quantum state is apparently high, the accuracy for classifying Werner states is rather low.
Similarly, the second subfigure of Fig.~\ref{fig:error-Fs} shows that the classification errors for generalized Werner states are  high.
Thus the models trained with features $F_2$ have poor generalization ability.
As a result, these classifiers have poor performance compared with traditional SDP method, as illustrated by Fig.~\ref{fig:errorBound-F2}.
Therefore, exploring high performance classifier with partial information is not trivial, and such a simple and crude way is impracticable.

To further explore high performance classifier with partial information, we  convert the state $\rho$ into a
canonical form $\rho_{0}$ by local unitaries, which preserves the steerability  of $\rho$.
As proved in \cite{Bowles}, the map given by,
\begin{eqnarray}\label{map}
\rho_{0}=(\mathrm{I}\otimes\rho_{B}^{1/2}) \rho (\mathrm{I}\otimes\rho_{B}^{1/2})
\end{eqnarray}
where $\rho_{B}=\mathrm{Tr}_{A}[\rho]$, preserves the steerability
of $\rho$. The interesting property of this map is that when applied
to an arbitrary state $\rho$, it can be realized by only local operation on Bob.

\begin{figure}
\begin{center}
\includegraphics[width=0.45\textwidth]{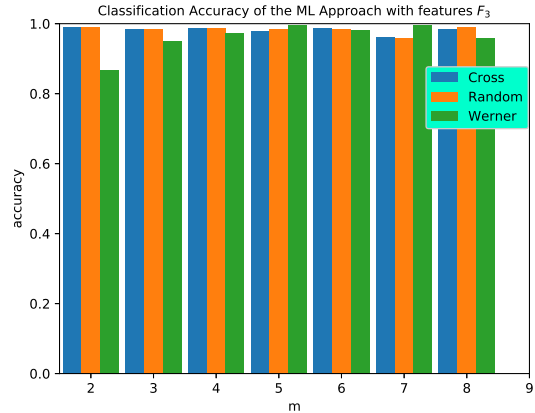}
 \caption{\label{fig:accuracy-F3}(Color online) \small{Classification accuracy of machine learning with partial information ($F_3$ features). The first column (blue) depicts the accuracy of cross validation; the second column (orange) depicts the classification accuracy on random states and the third column (green) depicts the classification accuracy on the Werner state.}}
\end{center}
\end{figure}

Similarly, we can extract the coefficients of the correlation terms of the resulting state $\rho_{0}$, $ \tau_{kl}$, to combine a real vector of $9$ numbers as features.
More precisely, we compute $\mathrm{Tr}[(\sigma_k\otimes \sigma_l)\rho_{0}]$ as features, denoted by $F_3$.
The classification accuracy of learned models for each $m$ is illustrated in Fig.~\ref{fig:accuracy-F3}.
In general, all the accuracies turn to higher and higher ($>0.95$) with the increase of $m$, which clearly shows that we get several well-trained learning machines.

Moreover, we observe the following similar phenomenon as using the full information features.
As illustrated in Fig.~\ref{fig:test8}, the green-star line is for $F_3$ features, it is shown that, the error drops very rapidly.
Hence, the variation tendency is identical to the theoretical prediction in general
(the more measurement settings, the more precise the prediction).

\begin{figure}[h]
\begin{center}
\includegraphics[width=0.45\textwidth]{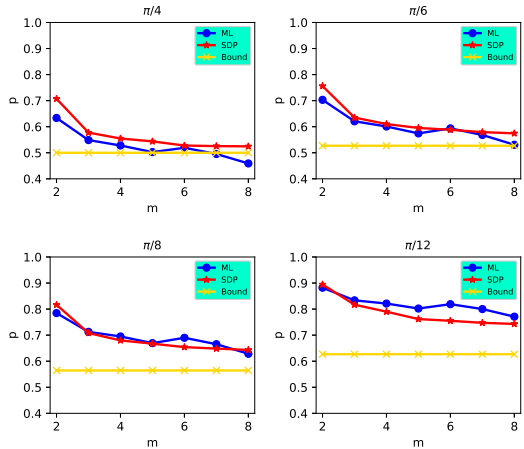}
\caption{\label{fig:errorBound-F3}(Color online) \small{The predictions of steerability for generalized Werner states by machine learning classifiers and SDP.
    The blue line is the result predicted by learned classifiers with $F_3$ features. Similarly, the red line is the result which predicted by SDP.
    And the orange line is the steerability bound from Alice to Bob.}}
 \end{center}
\end{figure}

Fig.~\ref{fig:errorBound-F3} illustrates the steerability bounds predicted by
machine learning classifiers and SDP with the angle $\xi=\{\frac{\pi}{4},\frac{\pi}{6},\frac{\pi}{8},\frac{\pi}{12}\}$ respectively.
Obviously, the learned classifiers outperforms the traditional SDP except for the states when $\xi=\frac{\pi}{12}$,
which is near the boundary of steerability. As the decrease of $\xi$, the prediction errors of both the learning classifiers and SDP increase.
As we mentioned for features $F_1$, it is a reasonable phenomenon since the predictions for the marginal states become harder and harder.
In Fig.~\ref{fig:error-Fs}, the third subfig shows the classification error for generalized Werner States (with different angles) for $F_3$ features.
Obviously, the error increases when the angle drops which coincides with the above analysis.
Generally, this classifier has better performance compared with traditional
SDP method. 

Note that in this scheme, for any unknown state, Alice and Bob only need to measure in three fixed measurement directions.
Therefore, it is more efficient than SDP in both physical measurement process and data processing.
 In particular, it should be very efficient for testing a large amount of arbitrary states in quantum information process, such as one-sided device-independent quantum key distribution (1SDI-QKD), channel discrimination  and teleamplification, \emph{etc}.

\begin{figure}
\begin{center}
\includegraphics[width=0.45\textwidth]{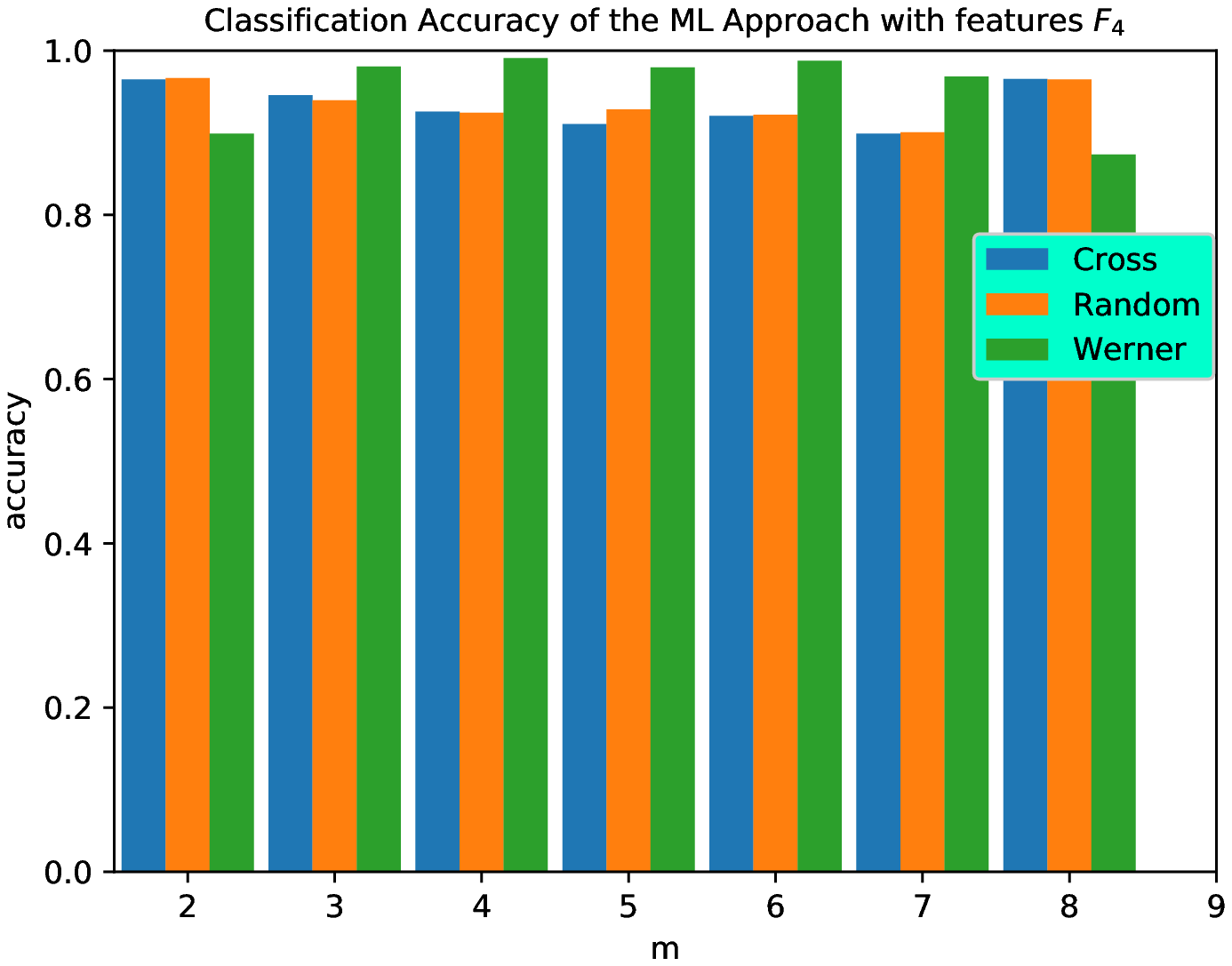}
 \caption{\label{fig:accuracy-F4}(Color online) \small{Classification accuracy of machine learning with partial information (F4 features). The first column (blue) depicts the accuracy of cross validation; the second column (orange) depicts the classification accuracy on random states and the third column (green) depicts the classification accuracy on the Werner state.}}
\end{center}
\end{figure}

\begin{figure}[h]
\begin{center}
\includegraphics[width=0.45\textwidth]{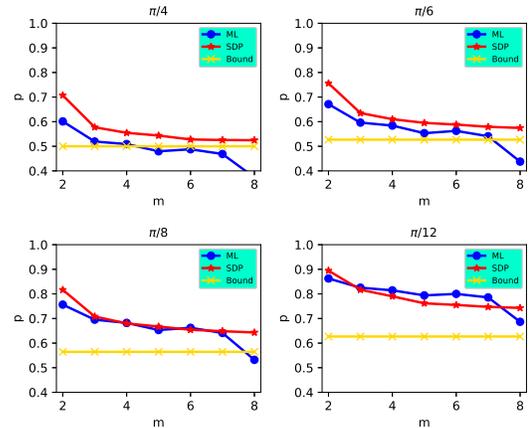}
 \caption{\label{fig:errorBound-F4}(Color online) \small{The predictions of steerability for generalized Werner states by learning classifiers and SDP. The blue line is the result predicted by learning classifiers with $F_4$ features. Similarly, the red line is the result predicted by SDP. And the orange line is the steerability bound from Alice to Bob.}}
 \end{center}
\end{figure}

To explore the performance of machine learning based quantum steering using even less information, according to the symmetry,
we dropped the coefficients of the correlation terms, $\sigma_y\otimes \sigma_x,\sigma_z\otimes \sigma_x,\sigma_z\otimes \sigma_y$ from $F_3$
and named the rest features as $F_4$. The training process was carried out as before.
As illustrated in Fig.~\ref{fig:accuracy-F4}, the classification accuracy of such classifiers on random states is acceptable but lower than with $F_3$ features.
Interestingly, it performs better on Werner states as illustrated in Fig.~\ref{fig:errorBound-F4} except for $m=8$.
The fact that the accuracy drops for $m=8$ may be caused by overfitting, as shown in Fig.~\ref{fig:accuracy-F4}.
Similarly, the predictions for the marginal states becomes harder and harder.
Even such classifiers with $F_4$ features perform worse than those with $F_3$ features,
they are still more effective than those with $F_2$ features.
Hence, correctly extracting partial information is very important for realizing high performance steering classifiers via machine learning.

\section{V. conclusion}
In this work, we have applied a method of machine learning to solve problems of quantum state classification in quantum
information science. Several reliable enhanced steerability classifiers
by combining supervised learning and the SDP method are achieved. 

At first, we build a high performance quantum steering classifier with the whole information, which are used to test some random unknown states and the generalized Werner states. The prediction performance of such learning classifier and SDP are completely analyzed and discussed. It clearly demonstrates the validity and efficiency of steering classification by machine learning.
Secondly, we investigate the possibility of constructing steering classifiers with partial information. It is shown that, correctly extracting partial information is very important for realizing high quality steering classifiers via machine learning. 
Finally, an efficient quantum steering detection scheme for arbitrary two-qubit states via machine learning is realized, where Alice and Bob only need to measure in three fixed measurement directions. 
%
It should be very efficient for testing the steerability of a large amount of arbitrary states in quantum information process, such as one-sided device-independent quantum key distribution (1SDI-QKD), channel discrimination  and teleamplification, \emph{etc}. 

\emph{Acknowledgement.-}C.L.R. is supported by National key research and development program (No. 2017YFA0305200), the Youth Innovation Promotion Association (CAS) (No. 2015317), the National Natural Science Foundation of China (No. 11605205), the Natural Science Foundation of Chongqing (No. cstc2015jcyjA00021, cstc2018jcyjAX0656), the Entrepreneurship and Innovation Support Program for Chongqing Overseas Returnees (No.cx017134), the fund of CAS Key Laboratory of Microscale Magnetic Resonance, and the fund of CAS Key Laboratory of Quantum Information.
And C.C. is supported by the National Natural Science Foundation of China (No. 11771421, 11471307, 61572024, 11671377), cstc2018jcyj-yszxX0002 of Chongqing,
and the Key Research Program of Frontier Sciences of CAS (QYZDB-SSW-SYS026). C. L. Ren and C. B. Chen contributed equally to this work.



\begin{appendices}



  \section{Appendix: The vector of features} For an arbitrary quantum state $\rho$, the four different features used in this work
  is summarized as below:
\begin{table}[htbp]
\begin{center}
\begin{tabular}{|c|c|}
\hline
\hline
\small{$\mathrm{F}_1$} & \small{$\rho_{ii}, i\in\{1,2,3\}$, the real and imaginary part of $\rho_{ij}, j>i$}  \\
\hline
\small{$\mathrm{F}_2$}  &\small{$\mathrm{Tr}[(\sigma_k\otimes \sigma_l)\rho]$, $\{k,l\}\in\{x,y,z\}$}\\
\hline
\small{$\mathrm{F}_3$}  &\small{$\rho\rightarrow\rho_0$, $\mathrm{Tr}[(\sigma_k\otimes \sigma_l)\rho_0]$, $\{k,l\}\in\{x,y,z\}$} \\
\hline
\small{$\mathrm{F}_4$}
&\small{$\mathrm{F}_3$ except for the terms of $\{\sigma_y\otimes \sigma_x,\sigma_z\otimes \sigma_x,\sigma_z\otimes \sigma_y\}$} \\
\hline
\hline
\end{tabular}
\end{center}
\end{table}

\end{appendices}

\end{document}